# Degenerate magnetic ground state and metastable state on trihexagonal Co-sublattices in $Co_3Sn_2(S,Se)_2$ single crystals

*Dong-Hwa Shin[*], Jin-Hyeon Jun, Sang-Eon Lee and Myung-Hwa Jung*

*Department of Physics, Sogang University, Seoul 04107, Republic of Korea*

**Abstract**

A trihexagonal lattice has been predicted to retain a degenerate magnetic state which enriches physical properties. Especially, $Co_3Sn_2S_2$, possessing trihexagonal Co-sublattices, has been observed to own topological quantum properties. Experimentally, $Co_3Sn_2S_2$ has been reported to have hidden magnetic phases due to magnetic anomalies. To clarify the hidden magnetic phase, we fabricated high-quality single crystals of $Co_3Sn_2S_{2-x}Se_x$ ($x$ = 0, 0.26, & 0.86). The Se-substitution is intended to broaden the distance between Co atoms. For each Se-composition, magnetizations of single-crystalline $Co_3Sn_2S_{2-x}Se_x$ implicate that the magnetic ground state consists of the out-of-plane ferromagnetism and the in-plane antiferromagnetism: Being degenerate. Meanwhile, along out-of-plane, remanent magnetizations of $Co_3Sn_2S_{2-x}Se_x$ have a first-order phase transition, so $Co_3Sn_2S_{2-x}Se_x$ owns an excited magnetic state, denoted as the metastable state. Consequently, we have provided semiclassical magnetic structures of the degenerate ground state and the metastable state via both magnetic symmetries and experimental constraints. Moreover, we have discovered mechanisms raising degeneracy and metastability.

[*] Authors to whom correspondence should be addressed: donghwashin728@gmail.com

# I. INTRODUCTION

A degenerate magnetic state, often called a quantum spin liquid, consists of multiple magnetic states which coexist at the same energy level [1–4]. The degenerate magnetic state is not a partial state but a macroscopic symmetric structure. Moreover, the degenerate magnetic state used to be described as a widely entangled spin fluctuation in terms of semiclassical physics. *i.e.,* the degenerate magnetic state seems to be a dynamic magnetic phase.

The degenerate magnetic state has been predicted to emerge easily on geometrically frustrated lattices − such as a pyrochlore lattice [5–8], a triangular lattice in spinel structure [4,9,10], and a trihexagonal lattice (Fig.1a) [9,11–15]. Experimentally, the degenerate magnetic state has been discovered on such lattices: For example, a 'spin ice' is a degenerate ground state satisfying tetrahedral symmetries in a pyrochlore, so the spin ice is semiclassically described as a 3-dimensional magnetic structure [5,6]; Furthermore, in Fe-based spinel $Fe_3O_4$, a spiral magnetic structure of $Fe_3O_4$ has a degeneracy caused by double-exchange interaction mechanism [16]. Furthermore, some degenerate magnetic states possess excited magnetic states – such as spin density wave in $Co_3V_2O_8$ [17], a helical magnetic state in $Mn_3Sn$ [18], and Kitaev paramagnetism in α-$RuCl_3$ [19].

We have focused on the trihexagonal lattice, predicted to retain topological quantum properties [9,11,13–15,20]. Experimentally, magnetic trihexagonal materials have been reported to own extraordinary physical characters: For instance, $Fe_3Sn_2$, containing trihexagonal Fe-sublattices, is a massive Dirac semimetal with noncollinear ferromagnetism [21,22]; (Fe/Co)Sn, possessing trihexagonal Fe/Co-sublattices, has Dirac cones and flat bands [23,24]; $Mn_3Sn$, retaining trihexagonal Mn-sublattices, is a Weyl semimetal with in-plane antiferromagnetic ground state and helical excited state [18,25]; Moreover, $Co_3Sn_2S_2$, including trihexagonal Co-sublattices, has been observed to hold topological quantum properties − such as a ferromagnetic Weyl phase [26–30], planar Hall effect [31,32], broken Ohm's law [33,34], a flat band [35,36], and giant skewness of anomalous Hall conductivity [37,38].

Experimentally, $Co_3Sn_2S_2$ has been reported to own hidden magnetic phases [39–43]. First, in AC susceptibility measurements with weak *H*-field, there is a peak below a Curie temperature ($T_C$), which indicates a new magnetic phase, differing from collinear ferromagnetism (FM) [40,42,43]. Second, in zero-field muon-spin rotation (zero-field μSR), there are both a primary muon-spin precession at 5 K and a secondary precession below $T_C$, which implies the existence of two distinct in-plane magnetic structures at 5 K and below $T_C$ [39]. Finally, exchange biased anomalous Hall effect at 2 K was observed, which means the coexistence of FM and antiferromagnetism (AFM) [43].

To clarify the hidden magnetic phases, we broadened the distance between Co atoms by substituting Se for S in $Co_3Sn_2S_2$. The Se-substitution has merit: $Co_3Sn_2Se_2$ has been expected to be in the same crystal phase of $Co_3Sn_2S_2$ and the magnetic Weyl phase [29]. However, bulk $Co_3Sn_2Se_2$ is unstable. Experimentally, for $x \geq 1$, bulk $Co_3Sn_2S_{2-x}Se_x$ cannot be fabricated [44]. As a result, we fabricated bulk single-crystal $Co_3Sn_2S_{2-x}Se_x$ of $0 \leq x < 1$.

## II. EXPERIMENTS

### A. Single crystal fabrication

To fabricate high quality single crystals, congruent mixtures (Co: Sn: S+Se = 3: 2: 2) were sealed in quartz tubes, were heated to 930 °C over 2 days, and were kept for 2 days before being slowly cooled to 700 °C over 7 days. Moreover, the stoichiometric mixtures were kept at 700 °C for 2 days to obtain translationally symmetric crystals. A crystal structure was determined by Cu Kα X-ray diffraction (XRD) whose X-ray wavelength is 1.5406 Å. As shown in Fig.1b, $Co_3Sn_2(S,Se)_2$ crystallizes as a rhombohedral structure: the space group $R\bar{3}m$ (no. 166). Fig.app1 is a picture of a $Co_3Sn_2(S,Se)_2$ single crystal. Its cleaved shining surface is an *ab*-plane. Define 'in-plane' (*ip*) as a direction parallel with *ab*-plane, and 'out-of-plane' (*op*) as *c*-axis. Compositions of single-crystalline samples were measured via a field emission electron probe microanalyzer (FE-EPMA). We have confirmed that all elements distribute uniformly. Denote Se-composition ($x$) given by

$$(\# \text{ of Co atoms}) : (\# \text{ of Se atoms}) = 3 : x \quad (1)$$

Co-composition is a reliable criterion, for S is evaporative. Fig.app2a & app2b are FE-EPMA data of $x = 0.26$ & $0.86$, respectively. Consequently, we have obtained three single crystals: $x = 0$, $0.26$, & $0.86$.

Fig.1c displays that crushed single crystals were characterized as $R\bar{3}m$ by powder X-ray diffraction (XRD). The powder XRD of $x = 0$ imply hexagonal lattice constants of $x = 0$: $a = 5.39(4)$ Å & $c = 13.2(9)$ Å which are similar to reported lattice constants [37]. Furthermore, red dashed lines in Fig.1c indicate that $x = 0.26$ & $0.86$ are also in the same phase. Nevertheless, the powder XRD of $x = 0.26$ & $0.86$ are slightly different to those of $x = 0$. Each peak of $x = 0.26$ & $0.86$ splits and translates to lower angles, which implicates the Se-substitution broadening lattice constants. Moreover, relative intensities are changed: For instance, the 40.94° peak of $x = 0$ is the highest intensity, yet those of $x = 0.26$ & $0.86$ are not. Such variations originate from which the radius of the Se atom is bigger than that of the S atom. The variations agree with the previous report [44]. In Appendix.A, we have verified that the Se-substitution satisfies translational symmetry by analyzing the peak variations.

## B. Magnetization measurement

Magnetic properties of $x$ = 0, 0.26, & 0.86 were measured by a superconducting quantum interference device-vibrating sample magnetometer (SQUID-VSM) in a magnetic field up to 7 T and temperature down to 2 K. $H$-fields are applied along $op$ and $ip$. Magnetizations depending on temperature were measured in three different methods: a zero-field-cooling (ZFC), a field-cooling (FC), and a remanent magnetization ($M_R$) measurement. The ZFC is a process of cooling without $H$-field and then of heating and measuring magnetic moments with 1 T. The FC is the process of cooling with 1 T and then of heating and measuring magnetic moments with 1 T. The $M_R$ measurement is the process of cooling with 7 T and then of heating and measuring magnetic moments without an $H$-field. Magnetizations along $op$ depending on magnetic field strength were measured at 2 K, $T_{com}$ −10 K, $T_{com}$ +10 K, $T_C$ −10 K, & 300 K. Magnetizations along $ip$ depending on magnetic field strength were measured at 2 K, $T_N$ −10 K, & 300 K.

## C. $M(T)$ measurement

In Fig.2a for $x$ = 0, $op$-ZFC/FC characterize magnetism of $x$ = 0 as FM with $T_C$ = 181.6 K, the minimum point of $\partial M/\partial T$ (the inset of $x$ = 0). In addition, the $T_C$= 181.6 K corresponds with reported data [37–43]. However, $ip$-ZFC/FC imply AFM with the Néel temperature $T_N$ = 177.2 K, the maximum point in $ip$-$M(T)$. As a result, $op$-FM and $ip$-AFM coexist at 2 K. Due to the coexistence, we shall denote the ground magnetic state of $x$ = 0 as a 'degenerate ground state'.

In Fig.2a for $x$ = 0, $op$-$M_R(T)$ shows a first-order phase transition whose transition temperature is denoted as '$T_{com}$'= 149.5 K, the minimum temperature at which $op$-$M_R$ compensates. The state in $T_{com}$ < $T$ < $T_C$ is not FM, for the FM has a second-order phase transition. As we mentioned in the introduction, the unknown state in $T_{com}$ < $T$ < $T_C$ has been observed indirectly under a weak $H$-field [40–43]. Hence, we shall denote the new degenerate excited state in $T_{com}$ < $T$ < $T_C$ as a

'metastable state'. The metastability is an adequate terminology, for the state emerges over the first-order phase transition and under weak $H$-field.

Likewise, $x = 0.26$ & $0.86$ also own the degenerate ground state ($T < T_{com}$) and the metastable state ($T_{com} < T < T_C$). Fig.2b shows the transition temperatures with respect to $x$. Transition temperatures of $x = 0$ are $T_C = 181.6$ K, $T_N = 177.2$ K, and $T_{com} = 149.5$ K; those of $x = 0.26$ are $T_C = 172.7$ K, $T_N = 166.9$ K, and $T_{com} = 134.7$ K; moreover, those of $x = 0.86$ are $T_C = 150.0$ K, $T_N = 141.2$ K, and $T_{com} = 90.2$ K. As $x$ increases, all transition temperatures decrease and $T_C - T_{com}$ increases, so the Se-substitution weakens ferromagnetic exchange interaction.

### D. *M(H)* measurement

As shown in Fig.3a for $x = 0$, we also measured $M(H)$. $op$-$M(H)$ were measured at 2 K, $T_{com} - 10$ K, $T_{com} + 10$ K, $T_C - 10$ K, & 300 K. Although, $op$-$M(H)$ imply FM at 2 K & $T_{com} - 10$ K, they do not at $T_{com} + 10$ K & $T_C - 10$ K. As shown in the inset for $x = 0$, $op$-$M(H)$ in $T_{com} < T < T_C$ possess neither remanent magnetizations nor coercivities. Hence, we shall denote the $op$-component of the metastable state as a 'weak $op$-AFM'. Meanwhile, $ip$-$M(H)$ of $x = 0$ were measured at 2 K, $T_N - 10$ K, & 300 K. At 2 K & $T_N - 10$ K, $ip$-$M(H)$ suggest $ip$-AFM under $T_N$: $ip$-$M(H)$ own neither remanent magnetizations nor coercivities, and the magnetizations of $ip$-$M(H)$ are smaller than those of $op$-$M(H)$. While, both $op$-$M(H)$ and $ip$-$M(H)$ at 300 K equal, which implicates paramagnetism. Likewise, $M(H)$ of $x = 0.26$ & $0.86$ indicate the same magnetic phases: the degenerate ground state under $T_{com}$, the metastable state in $T_{com} < T < T_C$, and the paramagnetism over $T_C$.

By the way, physical values derived from $M(H)$ suggest other information, such as sample quality (Appendix.B) and variation of exchange-interaction intensity. Even though the Se-substitution rarely affects $op$-$M_R$(2 K), ~1.02 $\mu_B$/f.u., the coercivity ($H_C$) at 2 K decreases from 5000 Oe at $x = 0$ to 2250 Oe at $x = 0.86$ (Fig.3b). Moreover, magnetocrystalline anisotropy energy density ($K_u$) increases from 0.90(0) MJ/m$^3$ at $x = 0$ to 0.96(4) MJ/m$^3$ at $x = 0.86$ (Fig.3b), where $K_u$ is obtained by

$$K_u = \mu_0 \int (M_{op,initial} - M_{ip}) dH \tag{2}$$

Note that $K_u$ of $x = 0$ is similar to the reported value [41]. Consequently, the Se-substitution reduces $H_C$ and enhances $K_u$, so the substitution weakens the ferromagnetic exchange interaction at 2 K.

### E. Semiclassical magnetic structures

#### 1. Summary for magnetic data

We have found out the magnetic phases of $Co_3Sn_2(S,Se)_2$: the degenerate ground state (*op*-FM + *ip*-AFM), the metastable state (weak *op*-AFM + *ip*-AFM). Also, we have verified that the Se-substitution weakens the ferromagnetic exchange interaction. Here, we shall figure out semiclassical, visualized magnetic structures of the degenerate ground state and the metastable state by using the magnetic symmetries in $Co_3Sn_2(S,Se)_2$ and their experimental constraints. Furthermore, we shall discover mechanisms of how degeneracy and metastability rise.

#### 2. All magnetic structures in $R\bar{3}m$

We shall use group theory to discover all possible magnetic structures. A group of magnetic symmetries is a subgroup of the space group $R\bar{3}m$ due to a relativistic duality between electric field and magnetic field: For example, if a space group has a 3-fold *ip*-symmetry, then 4-fold magnetic *ip*-symmetry cannot be allowed. Furthermore, the group of the magnetic symmetries is a normal subgroup of $R\bar{3}m$ (Appendix.C), so suffice to find all normal subgroups of $R\bar{3}m$. Since $R\bar{3}m$ consists of 3-fold *ip*-rotation (*r*), parity conjugation (*p*), and mirror conjugation (*m*), then $R\bar{3}m$ can be written as the following: For a group identity *e* and a cyclic group $C_n$ of order *n*,

$$R\bar{3}m = \langle r, p, m | r^3 = p^2 = m^2 = rpr^{-1}p^{-1} = (rm)^2 = e \rangle \cong C_6 \rtimes C_2 \tag{3}$$

*viz.*, $R\bar{3}m$ is group-isomorphic to the dihedral group of order 12. Since there are exactly 6 normal subgroups of the dihedral group of order 12, then there are 7 possible magnetic structures, including the dihedral group of order 12. Indeed, the trivial group $\langle e \rangle$ has no symmetry, so $\langle e \rangle$-structure is paramagnetism, the most degenerate magnetic state.

### 3. Mechanism raising degenerate ground state

Fig.4 is a Hasse diagram of $R\bar{3}m$ normal series and their corresponding magnetic structures. *viz.*, for groups $G$ and $G'$, $G \rightarrow G'$ implies that $G'$ is a normal subgroup of $G$. Note that $\langle r, p, m \rangle$-structure, satisfying all $R\bar{3}m$ symmetries, only consists of *ip*-structure, for *m*-symmetry prohibits *op*-components (Appendix.D). However, the degenerate ground state and the metastable state have *op*-structure, so *m*-symmetry has to be broken at 2 K. In fact, an *m*-breaking mechanism is antisymmetric superexchange interaction mediated by S/Se atoms. In other words, since S/Se sites are not centers of *p*-conjugation, then the antisymmetric term of superexchange interaction cannot be canceled out. [45] Thus, antisymmetric superexchange interaction raises *op*-FM [11]. Indeed, o*p*-FM satisfies *r*-symmetry and *p*-symmetry, so the degenerate ground state is $\langle r, p \rangle$-structure.

### 4. Mechanism raising metastable state

On the other hand, the metastable state is a substructure of the degenerate ground state, either $\langle p \rangle$-structure or $\langle r \rangle$-structure. Using a proof of contradiction, we shall verify that the metastable state violates *r*-symmetry. Were the metastable state $\langle r \rangle$-structure, the state would violate *p*-symmetry. Due to *p*-violation, $\langle r \rangle$-structure possesses an expanded magnetic unit cell along the *c*-axis (Fig.app5), called a spin density wave. However, neutron scattering measurements of $Co_3Sn_2S_2$ indicate no spin density wave [39,46], which contradicts. As a result, the metastable state is not $\langle r \rangle$-structure but $\langle p \rangle$-structure, an unexpanded *op*-AFM.

There is an *r*-violating mechanism which is an expansion of distance ($a/2$) between Co atoms. In detail, thermal expansion causes $a/2$-expansion in $Co_3Sn$ layers and brings out an antiferromagnetic exchange interaction, explained by Ruderman-Kittel-Kasuya-Yosida (RKKY) interaction between Co *d*-orbitals [47–49]. To substantiate the mechanism, we broadened the distance between Co atoms by the Se-substitution. Recall that $T_{com}$ decreases as $x$ increases, which indicates the Se-substitution enhances the antiferromagnetic exchange interaction.

### 5. Semiclassical magnetic structure of metastable state

Based on *p*-symmetry and *r*-violation, we have discovered a semiclassical structure of the metastable state along with *op* & *ip*. First, over $T_{com}$, the antiferromagnetic exchange interaction overcomes the ferromagnetic exchange interaction, so the *p*-symmetric unexpanded *op*-AFM emerges. The unexpanded *op*-AFM consists of upward moments, downward moments, and zero moments. The zero moments are the degeneracy of both the upward moments and the downward moments. In addition, the unexpanded *op*-AFM seems to be the semiclassical expression of Wannier's degenerate *op*-AFM [4].

On the other hand, the antiferromagnetic exchange interaction also affects *ip*-AFM of the metastable state. Note that *r*-symmetry only allows *ip*-AFM to possess a positive winding number which is a topological invariant in (pseudo)vector field $M_{ip}$, given by

$$\text{(winding number)} = \frac{1}{2\pi} \oint M_{ip} \tag{4}$$

However, the antiferromagnetic exchange interaction, *r*-violation, causes a negative winding number that a secondary *ip*-AFM retains. There is experimental evidence for the negative *ip*-AFM in the metastable state. First, the secondary precession of zero-field μSR indicates a new *ip*-AFM, differing from the positive *ip*-AFM of the degenerate ground state [39]. Note that the secondary precession overcomes the primary precession near $T_{com}$. Second, another evidence of negative *ip*-AFM is a giant skewness of Hall current ($=\sigma_{xy}/\sigma_{xx}$) of $Co_3Sn_2S_2$ near $T_{com}$=150 K [37]. Since there are two *ip*-AFM of opposite winding numbers, neither are dominant near $T_{com}$, then a difference between opposite topological invariants maximizes. As a result, because of topological conservation law, the skewness also maximizes near $T_{com}$ [50].

### III. CONCLUSION

As shown in Fig.5, we have visualized semiclassically the degenerate magnetic state and the metastable state of $Co_3Sn_2(S,Se)_2$. Although there are microscopic, partial magnetic fluctuations, we may choose macroscopic, symmetric magnetic backgrounds, guaranteed by the axioms of quantum physics. Our study implicates two scientific significances. First, both good-quality single crystals and their magnetic data are enough to find magnetic structures. Second, we show that $Co_3Sn_2(S,Se)_2$ is an excellent platform to research electromagnetic properties in degenerate magnetic states. For instance, in a 3-dimensional magnetic structure, a conducting electron is affected by the Aharonov-Bohm effect, described as a virtual magnetic field (or a real-space Berry curvature). Indeed, the virtual magnetic field boosts anomalous Hall conductivity. Likewise, Co-based shandite, $Co_3Sn_2(S,Se)_2$, is an attractive material group because of its 3-dimensional magnetic structure or its magnetic degeneracy.

Magnetizations of the degenerate magnetic states seem to be complicated, so studying the degenerate magnetic states used to demand various measurements − such as AC susceptibility, magnetic specific heat measurement, μSR, neutron diffraction, and neutron spectrometer [2]. Nevertheless, magnetization suggests two essential clues to finding magnetic structures: (1) the existence of degeneracy and (2) constraint for the number of possible magnetic structures. These are our new paradigm to analyze magnetizations of the degenerate magnetic states.

## APPENDIX A

We shall prove that substituted samples are said to be translationally symmetric. In Fig.app3, the ($20\bar{2}4$) peak of $x = 0$ is isolated and non-overlapped. While, the peaks ($20\bar{2}4$) of $x = 0.26$ & $0.86$ are shifted to lower angle and split into two convoluted peaks. We have deconvoluted each of the peaks by using the Pearson type VII distribution $p(x)$.

$$p(x) = \frac{1}{\alpha B\left(m - \frac{1}{2}, \frac{1}{2}\right)} \left[1 + \left(\frac{x - \lambda}{\alpha}\right)^2\right]^{-m} \tag{a1}$$

For each $x$, the lower peak (red) is the ($20\bar{2}4$) reflection of $Co_3Sn_2Se_2$, and the higher peak (blue) is the ($20\bar{2}4$) reflection of $Co_3Sn_2S_2$, for the mean distance of the red peak is broader than that of the blue peak. First of all, recall that emerging XRD reflection requires translational symmetry. Consequently, the existence of red & blue peaks indicates the translational symmetry of $x = 0.26$ & $0.86$.

## APPENDIX B

Sample quality is how translationally symmetric a crystal is. The sample quality is essential to analyze magnetic properties: If the sample quality is poor, then we cannot determine whether degenerate magnetism or amorphous magnetism. Here, there are three criteria for sample quality derived from $M(H)$: (1) the linearity of an initial curve, (2) a coercivity ($H_C$), and (3) a squareness of $M(H)$, $M(0\ T)/M(7\ T)$.

Fig.app4 displays $op$-$M(2\ K, H)$ of $x = 0, 0.26,$ & $0.86$ (left) and $M(2\ K, H)$ of a pristine powder (right). The pristine powder sample lacks translational symmetry, so it represents a poor-quality sample. First, the initial curves of $op$-$M(2\ K, H)$ of $x = 0, 0.26,$ & $0.86$ are linear, which indicates a few magnetic domains in samples. In other words, were there a lot of magnetic domains in samples, then the initial curves would be non-linear, well known for Rayleigh law. Second, the coercivity $H_C$ is also a standard for the sample quality. $x = 0$ has $H_C = 5000$ Oe at 2 K, yet it is much larger than reported $H_C$ (500~800 Oe)[1–7]. In fact, our samples have better quality, since the pristine powder has

$H_C$ = 200 Oe. Finally, a squareness ratio of hysteresis, $M(0\,T)/M(7\,T)$, is also one of the criteria for the sample quality. All single-crystalline samples have a squareness ratio of ~98%, yet the pristine powder has a 58% squareness ratio given by

$$\frac{M(0\,T)}{M(7\,T)} \times 100\,\% = \frac{0.55\,\mu_B/\text{f.u.}}{0.95\,\mu_B/\text{f.u.}} \times 100\,\% = 58\,\% \tag{a2}$$

Hence, the higher the squareness is, the better the sample quality is. Consequently, it is guaranteed that our samples have good quality. *i.e.*, $x = 0$, 0.26, & 0.86 are much translationally symmetric.

# APPENDIX C

Here, we shall prove that a group of magnetic symmetries is a normal subgroup of a space group. Since we have verified that the group of magnetic symmetries is a subgroup of the space group, suffice to show that it is also the normal subgroup.

**Definition:** A subgroup $N$ of a group $G$ is said to be a 'normal subgroup' of $G$ provided that, for any $g$ in $G$ and $n$ in $N$, there is $n'$ in $N$ given by

$$n'g = gn \tag{a3}$$

**Definition:** For a magnetization $M$ and transformation $g$ in a space group $G$, $M$ is said to be $g$-symmetric if $M$ is invariant under a $g$-conjugation:

$$\det(g)\, g \circ M \circ g^{-1} = M, \tag{a4}$$

where the determinant ($\pm 1$) is required due to the pseudovector field $M$.

**Theorem:** Given a space group $G$ and its subgroup $N$, let a magnetization $M$ be an $N$-structure, satisfying all conjugations in $N$. Then $N$ is a normal group.

*pf)* Take any transformation *g* in *G*. Since *g*-conjugated *M* is also *N*-structure, the following holds:

$$\det(ng)\, ng \circ M \circ (ng)^{-1} = \det(n)\det(g)\, n \circ (g \circ M \circ g^{-1}) \circ n^{-1} = \det(g)\, g \circ M \circ g^{-1} \quad (a5)$$

Since *M* is *N*-structure, we may choose *n'* in *N* satisfying det(*n'*)=det(*n*). Hence,

$$\det(g)\, g \circ M \circ g^{-1} = \det(g)\det(n')\, g \circ (n' \circ M \circ n'^{-1}) \circ g^{-1} = \det(gn')\, gn' \circ M \circ (gn')^{-1} \quad (a6)$$

As a result,

$$\det(ng)\, ng \circ M \circ (ng)^{-1} = \det(gn')\, gn' \circ M \circ (gn')^{-1} \quad (a7)$$

*i.e., ng=gn'.*∎

## APPENDIX D

**Proposition:** *m*-symmetry prohibits *op*-components.

*proof)* Let *m*-symmetry holds. Observe that each Co atom contains at least one mirror plane, so that

$$M \circ m^{-1} = M \quad (a8)$$

Also, all mirror planes contain the *c*-axis. Given $M^c \hat{c}$, *c*-axis component of magnetization, the mirror conjugation is written as follow:

$$M^c \hat{c} = \det(m)\, m \circ (M^c \hat{c}) \circ m^{-1} = (-1) m \circ (M^c \hat{c}) = (-1)(M^c \hat{c}) \quad (a9)$$

∴ $M^c = 0$ ∎

**Acknowledgements**

This work was supported by the National Research Foundation of Korea (NRF) grant funded by the Korean government (MEST) (Nos. 2020R1A2C3008044 and 2016M3A7B4910400).

Figures

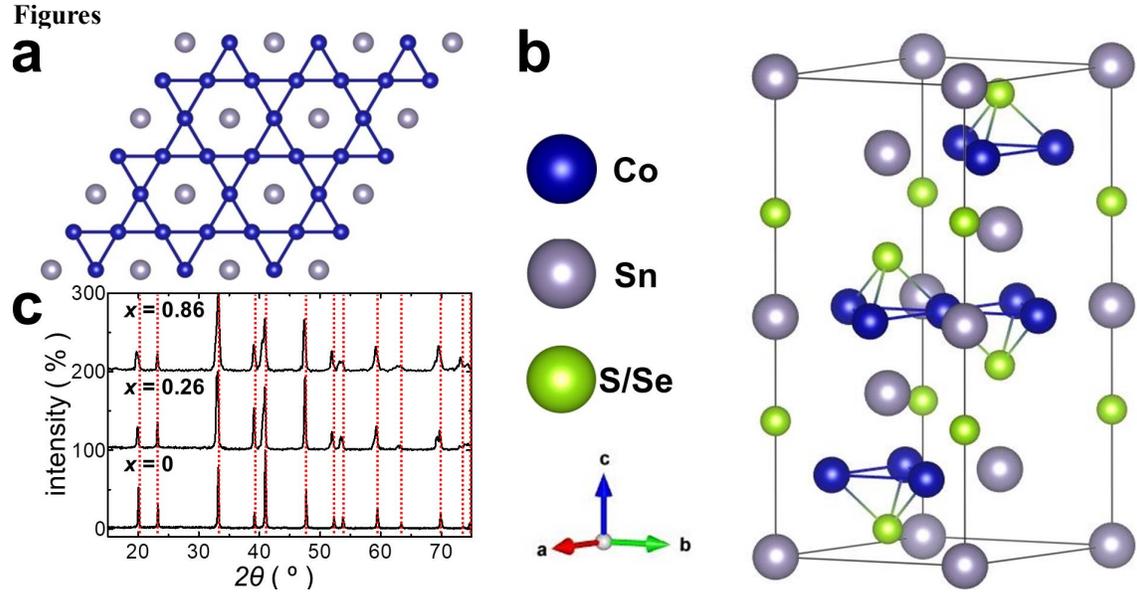

**FIG. 1.** Crystal structure of $Co_3Sn_2(S,Se)_2$ **(a)** Trihexagonal lattice, well-known as a kagomé lattice, consists of corner-sharing equilateral triangles. **(b)** The crystal structure ($R\bar{3}m$, no.166) of $Co_3Sn_2(S,Se)_2$ in terms of the hexagonal Bravais lattice. $Co_3Sn_2(S,Se)_2$ contains 2-dimensional $Co_3Sn$ layers sandwiched among S and Se atoms: trihexagonal Co-sublattices. The crystal structure satisfies three symmetries: 3-fold in-plane rotational symmetry, parity symmetry, and mirror symmetry. **(c)** Powder XRD of crushed $Co_3Sn_2S_{2-x}Se_x$ single crystals, $x$ = 0, 0.26 & 0.86. Red dashed lines are peak positions of powder XRD for $x$ = 0. The samples of $x$ = 0.26 & 0.86 have similar peak positions with respect to $x$ = 0.

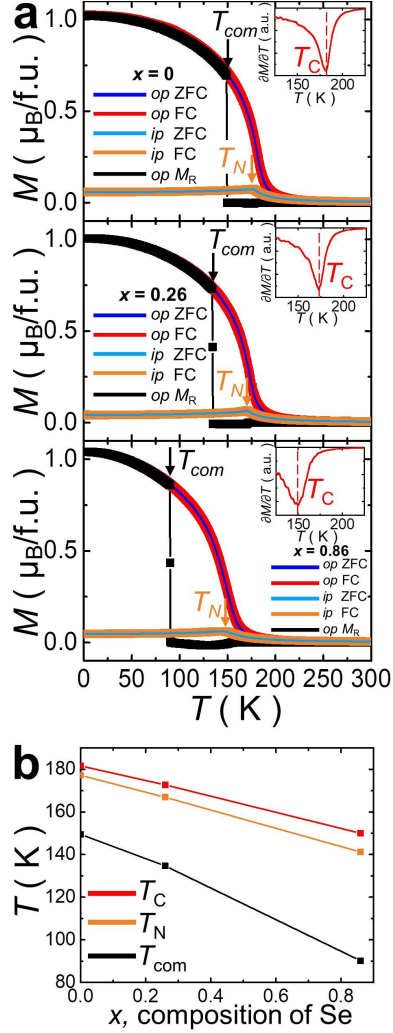

**FIG. 2.** Magnetization depending on temperature of $Co_3Sn_2S_{2-x}Se_x$ **(a)** For $x = 0$, 0.26, & 0.86, $M(T)$ of $Co_3Sn_2S_{2-x}Se_x$. *op*-ZFC (blue) and *op*-FC (red) with 1 T field have the same values, and they indicate ferromagnetism. The insets are the partial derivatives of *op*-FC with respect to $T$, so the minimum points of the insets are the Curie temperatures $T_C$: 181.6 K, 172.7 K, & 150.0 K of $x = 0$, 0.26 & 0.86, respectively. *ip*-ZFC (celeste) and *ip*-FC (orange) with 1 T field also have the same values, and they imply antiferromegnetism. The maximum points of *ip*-$M(T)$ are said to be Néel temperatures $T_N$: 177.2 K, 166.9 K, & 141.2 K of $x = 0$, 0.26 & 0.86, respectively. *op*-$M_R$ (black) are remanent magnetization depending on temperature. Each *op*-$M_R$ has a first-order phase transition $T_{com}$: 149.5 K, 134.7 K, & 90.2 K of $x = 0$, 0.26 & 0.86, respectively. **(b)** As $x$ increases, transition temperatures, $T_C$, $T_N$ & $T_{com}$ decease.

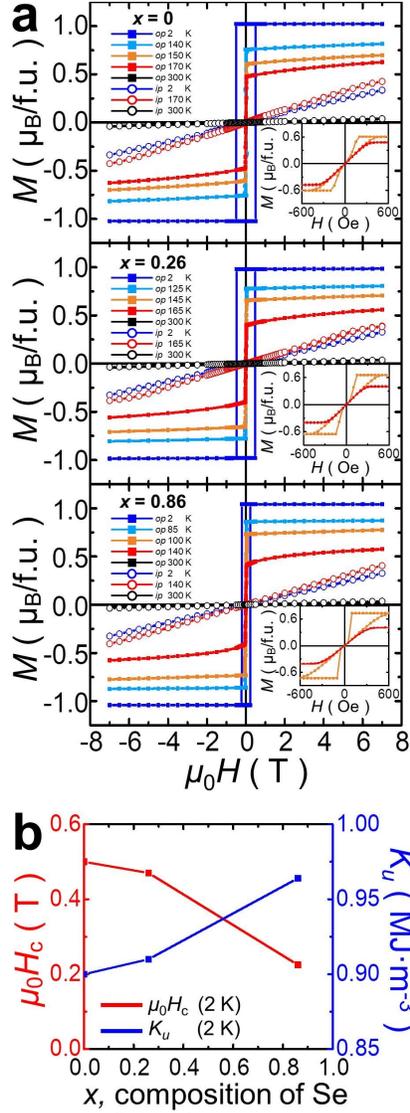

**FIG. 3.** Magnetization depending on $H$-field of $Co_3Sn_2S_{2-x}Se_x$ **(a)** For $x = 0$, $0.26$, & $0.86$, $M(H)$ of $Co_3Sn_2S_{2-x}Se_x$. For each $x$, *op-M(H)* at $T=2$ K (blue) & $T_{com}-10$ K (celeste) indicate FM, yet *op-M(H)* at $T_{com}+10$ K (orange) & $T_C-10$ K (red) do not. Each of the insets is *op-M(H)* at $T_{com}+10$ K & $T_C-10$ K up to $\pm$ 600 Oe. They possess neither remanent magnetizations nor coercivities. For each $x$, *ip-M(H)* at $T=2$ K (blue) and $T_C-10$ K (red) implicate AFM below $T_N$. For every $x$, both *op-M(H)* and *ip-M(H)* at 300 K (black) have the same values, so they imply paramagnetism. **(b)** Coercivities $\mu_0H_C$ (red) and magneto-crystalline anisotropy energy densities $K_u$ (blue) at 2 K. $x = 0$, $0.26$ & $0.86$ have $H_C$: 5000 Oe, 4700 Oe, & 22550 Oe, respectively. $x = 0$, $0.26$ & $0.86$ have $K_u$: 9.0(0) MJ·m$^{-3}$, 9.1(0) MJ·m$^{-3}$, & 9.6(4) MJ·m$^{-3}$, respectively. As $x$ increases, $\mu_0H_C$ decreases, while $K_u$ increases.

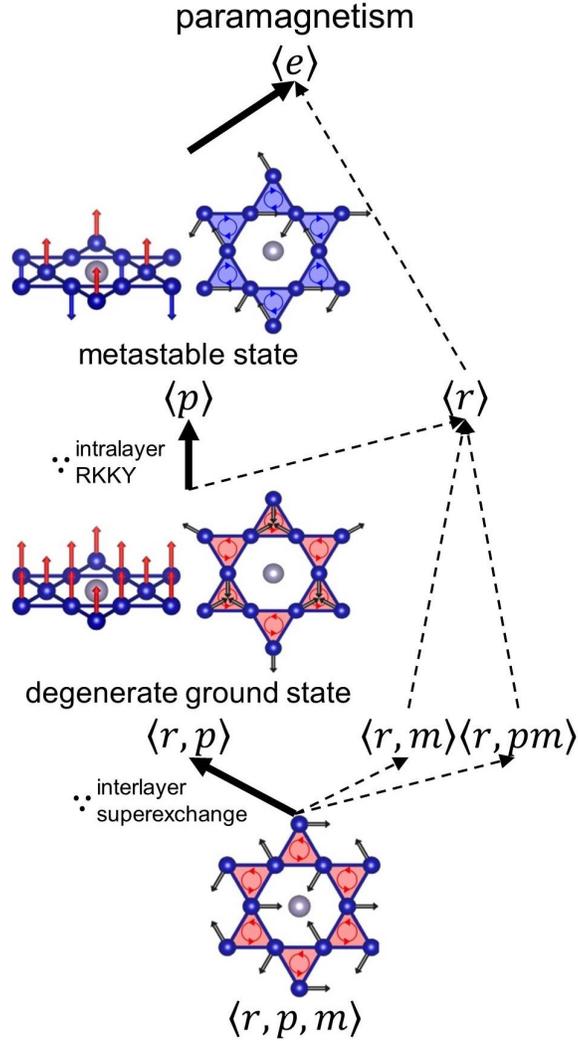

**FIG. 4.** Hasse diagram of $R\bar{3}m$ normal series. For groups $G$ and $G'$, $G \rightarrow G'$ is defined as $G'$ is a normal subgroup of $G$: $G \rhd G'$. $\langle r, p, m \rangle$-structure satisfying all symmetries of $R\bar{3}m$ only consists of *ip*-AFM with a positive winding number. Hence, neither the degenerate ground state nor the metastable state is $\langle r, p, m \rangle$-structure. Meanwhile, *op*-FM satisfies *r*-symmetry and *p*-symmetry, so the degenerate ground state is $\langle r, p \rangle$-structure. An unexpended *op*-AFM holds only *p*-symmetry. The unexpended *op*-AFM is composed of upward moments (red), downward moments (blue), and zero moments. As a result, the metastable state is $\langle p \rangle$-structure. Due to *r*-violation, *ip*-AFM of the metastable state owns a negative winding number. At last, paramagnetism holds no symmetry, so it is $\langle e \rangle$-structure.

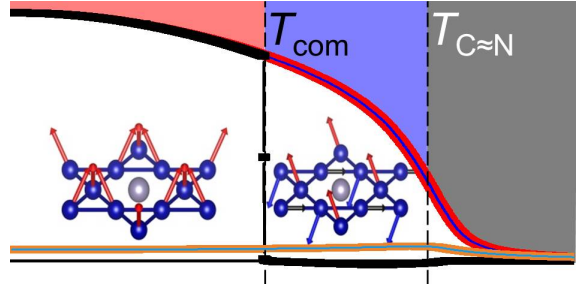

**FIG. 5.** Magnetic phase diagram of $Co_3Sn_2(S,Se)_2$. Semiclassical 3-dimensional magnetic structures of the degenerate ground state (*op*-FM + positive *ip*-AFM) and the metastable state (weak *op*-AFM + negative *ip*-AFM).

**FIG. app1.** A single crystal of $Co_3Sn_2(S,Se)_2$. The shining cleaved surface is an *ab*-plane.

**FIG. app2.** FE-EPMA mapping data of **(a)** $x = 0.26$ & **(b)** $x = 0.86$

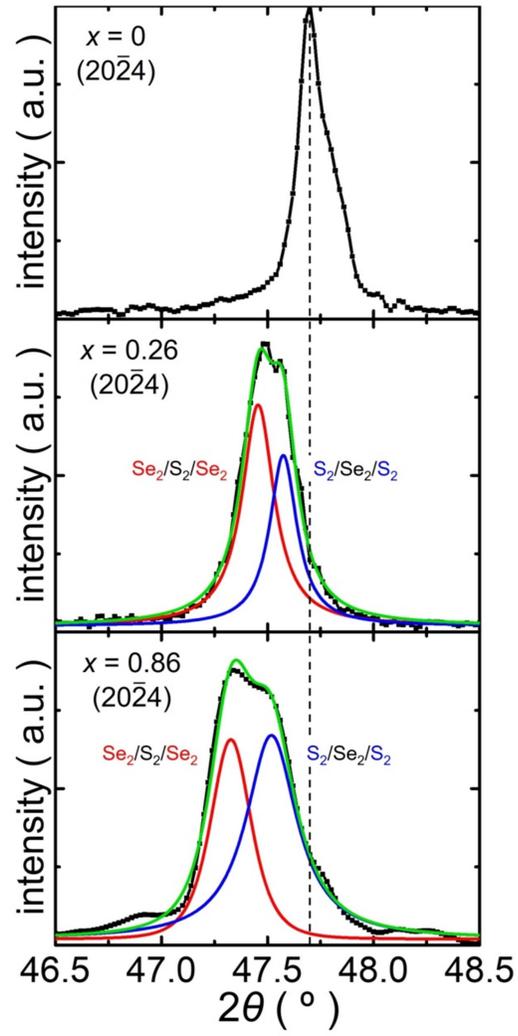

**FIG. app3.** (20$\bar{2}$4) peaks of $x$ = 0, 0.26, & 0.86. For each $x$=0.26 & 0.86, (20$\bar{2}$4) peaks of are deconvoluted into a lower peak (red) and a higher peak (blue). For each $x$, the lower peak is the (20$\bar{2}$4) reflection of $Co_3Sn_2Se_2$, and the higher peak is the (20$\bar{2}$4) reflection of $Co_3Sn_2S_2$.

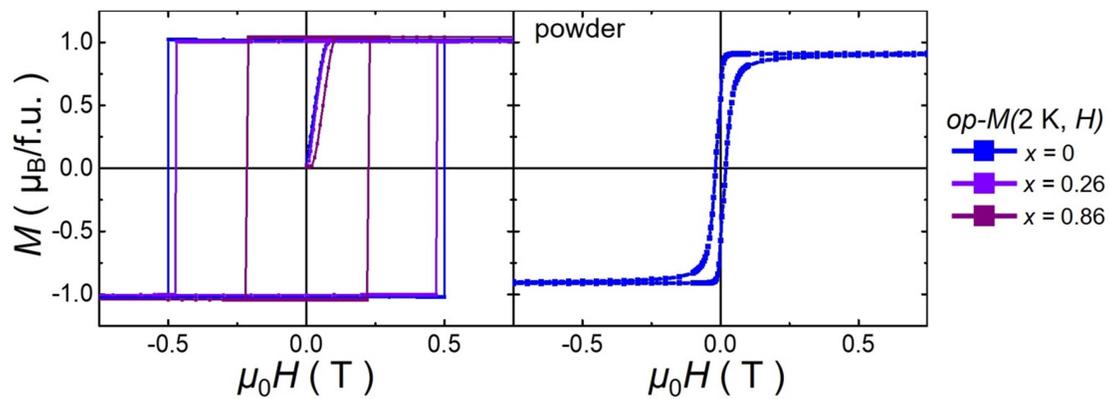

**FIG. app4.** At 2 K, *op-M(H)* of *x* = 0, 0.26, & 0.86 (left) and *M(H)* of pristine powder (right). Initial curves of *M(H)* are upward curves, starting at the origin (*H,M*) = (0,0).

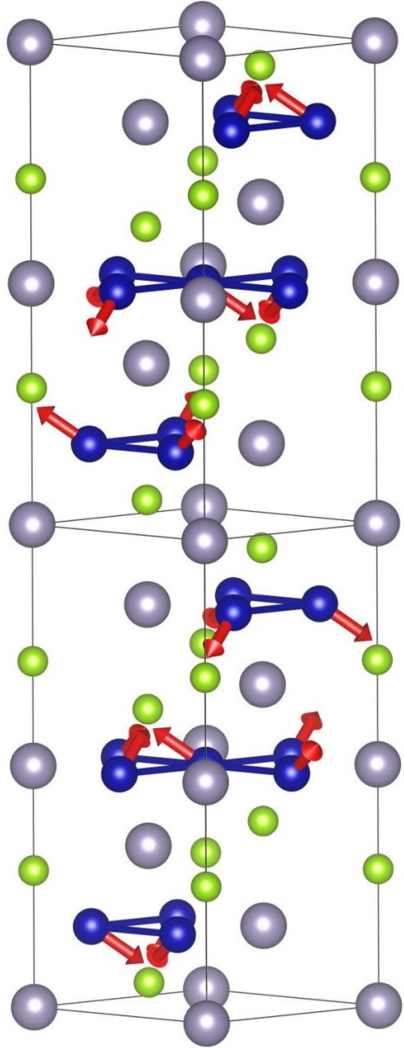

**FIG. app5.** ⟨r⟩-structure satisfies weak *op*-AFM and positive *ip*-AFM. Due to *p*-violation, magnetic moments of *p*-conjugation pairs are distinct. Hence the magnetic unit cell of ⟨r⟩-structure is expanded along the *c*-axis: ⟨r⟩-structure is a spin density wave. ⟨r⟩-structure only possesses a positive winding number, which disagrees with an existence of a secondary *ip*-AFM